\documentclass[reqno]{amsart}
\usepackage{amsfonts}
\usepackage{amsthm}
\usepackage{amssymb}
\usepackage{dsfont}
\usepackage{amscd}
\usepackage{color}
\usepackage{tikz}
 \usetikzlibrary{calc,decorations.markings,matrix,decorations.pathmorphing}

\newtheorem{theorem}{Theorem}[section]

\newtheorem{lemma}[theorem]{Lemma}
\newtheorem{prop}[theorem]{Proposition}
\theoremstyle{definition}

\theoremstyle{remark}

\numberwithin{equation}{section}

\def\be{\begin{equation}}
\def\ee{\end{equation}}
\def\ba{\begin{eqnarray*}}
\def\ea{\end{eqnarray*}}
\def\bae{\begin{eqnarray}}
\def\eae{\end{eqnarray}}
\def\bc{\begin{center}}
\def\ec{\end{center}}

\begin{document}

\title[Asymptotics of finite system Lyapunov exponents ]{Asymptotics of finite system Lyapunov exponents for some random matrix ensembles}
\author{Peter J. Forrester} \address{Department of Mathematics and Statistics, The University of Melbourne, Victoria 3010, Australia}

\begin{abstract}
For products $P_N$ of $N$ random matrices of size $d \times d$, there is a natural notion of finite $N$ Lyapunov
exponents $\{\mu_i\}_{i=1}^d$. In the case of standard Gaussian random matrices with real, complex or real quaternion elements, and extended to the general variance case for $\mu_1$, methods known for the computation of
$\lim_{N \to \infty} \langle \mu_i \rangle$ are used to compute the large $N$ form of the variances of the exponents.
Analogous calculations are performed in the case that the matrices making up $P_N$ are products of sub-blocks of
random unitary matrices with Haar measure.  Furthermore, we make some remarks relating to the
coincidence of the Lyapunov exponents and the stability exponents relating to the eigenvalues of $P_N$.
\end{abstract}

\maketitle
\section{Introduction}
Let $A_i$, $i=1,\dots,N$ be $d \times d$ random matrices independently chosen from the same ensemble, and form the product
\begin{equation}\label{1}
P_N = A_N A_{N-1} \cdots A_1.
\end{equation}
Under the condition that the second moment of the diagonal entries of $A_i^\dagger A_i$ are finite, the multiplicative ergodic theorem of 
Oseledec \cite{Os68,Ra79} tells us that the limiting positive definite matrix
\begin{equation}\label{2}
V_d := \lim_{N \to \infty} (P_N^\dagger P_N)^{1/(2N)}
\end{equation}
is well defined. In particular $V_d$ has eigenvalues
\begin{equation}\label{3}
e^{\mu_1} \ge e^{\mu_2} \ge \cdots \ge e^{\mu_d}
\end{equation}
with $\{\mu_i\}$ referred to as the Lyapunov exponents.

The recent study \cite{ABK14}, building on \cite{AKW13}, gave the asymptotic form of the eigenvalue distribution of $(P_N^\dagger P_N)^{1/(2N)}$ in the case that
each $A_i$ was chosen from the ensemble of $d \times d$ standard complex Gaussian matrices. For large $N$, and with the
eigenvalues of $P_N^\dagger P_N$, $\{x_i\}$ say, parametrised by writing $x_i = e^{2N y_i}$, this was shown to be
given by
\begin{equation}\label{4}
{1 \over N!} {\rm Sym} \, \prod_{j=1}^d {1 \over (2 \pi \sigma_i^2)^{1/2}} e^{- (y_i - \mu_i)^2/2 \sigma_i^2}
\end{equation}
where, with $\Psi(x)$ denoting the digamma function
\begin{equation}\label{4.1}
\mu_i = {1 \over 2} \Psi(d - i + 1), \qquad \sigma_i^2 = {1 \over 4N} \Psi'(d-i+1).
\end{equation}
In the $N \to \infty$ limit, it follows from this and the definitions in the first paragraph that the Lyapunov exponents are given
by ${1 \over 2} \Psi(d - i + 1)$, $(i=1,\dots,d)$, which is a known result \cite{Fo13}. The value of $\sigma_i^2$ for $i=1$ agrees with
another known result \cite{CN84}. One of the aims of the present paper is to show how $\{ \sigma_i^2 \}$ for products of Gaussian random
matrices can be computed by analogous calculations leading to $\{\mu_i\}$ \cite{Ne86a,Fo13,Ka14}. We do this
in Section \ref{S2.2}, and we point out there that the mean and variance of the finite $N$ Lyapunov exponents for $P_N$ consisting as a product of Gaussian
and inverse Gaussian matrices follows as a corollary.

The formulation used  to perform the computations in Section \ref{S2.2}, which itself is presented in Section \ref{S2.1}, allows for the computation of the variances associated with the distribution of $\{y_i\}$ for large but finite $N$ to also be computed in a number of other cases. One is for random matrices $A_i = 
\Sigma^{1/2} G_i$, where $\Sigma$ is a fixed $d \times d$ positive definite matrix and $G_i$ a standard complex Gaussian random matrix. This is carried out in Section \ref{S2.3}. In Section \ref{S2.4} we generalise 
$A_i = 
\Sigma^{1/2} G_i$ to the
case that $G_i$ is a standard real, or standard real quaternion, Gaussian random matrix, but for the variance corresponding to the largest Lyapunov exponent only. The Lyapanov exponents and associated variances are
computed for products of sub-blocks of random unitary matrices with real, complex or real quaternion entries and chosen with Haar measure in Section \ref{S3}. We point out that the formulas for the Lyapunov exponents and
the associated variances are
analogous to those for products involving Gaussian random matrices of two different sizes in equal proportion,
except for a sign.

The paper  \cite{ABK14} also undertook a study of the eigenvalues of the matrix $P_N$, $\{z_k\}$ say, parameterised by writing
$z_k = e^{N \lambda_k + i \theta_k}$. It was found that for large $N$ the joint distribution of $\{\lambda_k, \theta_k\}$ is given
by (\ref{4}), with each $y_i$ replaced by $\lambda_i$, and multiplied by $1/(2 \pi)^d$. Thus the angular dependence is uniform,
while the asymptotic distribution of $\{ \lambda_i \}$ is the same as that for $\{\mu_i\}$. On this latter point, following \cite{GSO87} as reported in the reference book
\cite[pg.~21]{CPV93}, we argue that provided $P_N$ can be diagonalised, the Lyaponov exponents can be computed from either the eigenvalues
of $P_N^\dagger P_N$ or the eigenvalues of $P_N$ itself. This point is made in Section \ref{S4.2}.

\section{Computation of variances for Gaussian ensembles}
\subsection{Known results}\label{S2.1}
While for a general ensemble of $d \times d$ random matrices the Lyapunov spectrum is difficult to compute, there is a class of probability densities
on the space of matrices --- referred to as isotropic --- for which the computation becomes relatively simple \cite{CN84}. The defining feature of an
isotropic matrix distribution on $\{A_i\}$ is that for a particular vector $\vec{x}$, $A_i \vec{x} /|\vec{x}|$ has distribution independent of $\vec{x}$.
Examples are when $A_i = \Sigma^{1/2} G_i^{(\beta,d)}{}$, where $\Sigma$ is a fixed $d \times d$ positive definite matrix and $G_i^{(\beta,d)}{}$ is
a $d \times d$ standard Gaussian matrix with real $(\beta = 1)$, complex $(\beta = 2)$ or real quaternion $(\beta = 4)$ entries. In the latter case,
these entries are themselves $2 \times 2$ matrices with complex entries, and the eigenvalues are calculated from the corresponding complex matrix
which has a doubly degenerate spectrum;
see e.g.~\cite[\S 1.3.2]{Fo10}.

The utility of an isotropic matrix distribution on $\{A_i\}$ can be seen by revising the formalism devised in \cite{Os68,Ra79} for the computation of the
Lyapunov spectrum.  The main formula specifies the partial sum of Lyapunov exponents according to
\begin{equation}\label{7.1}
\mu_1 + \cdots + \mu_k = \sup \lim_{N \to \infty} {1 \over N}  \log {\rm Vol}_k \{y_1(N),\dots,y_k(N) \} \qquad (k=1,\dots,d),
\end{equation}
where $y_j(N) := P_N y_j(0)$, the sup operation is over all sets of linearly independent vectors
$\{y_1(0),\dots,y_k(0) \}$ and Vol${}_k$ refers to the volume of the parallelogram generated by the given set of $k$ vectors.
To quantify the latter, with 
$$
B_N := [ y_1(N) \: y_2(N) \: \cdots  \: y_k(N)],
$$
so that $B_N$ is the $d \times k$ matrix with its columns given by the $k$ vectors $\{y_j(N)\}$, we have
\begin{equation}\label{7.2}
 {\rm Vol}_k \{y_1(N),\dots,y_k(N) \} = \det (B_N^\dagger B_N)^{1/2} = \det (B_0^\dagger P_N^\dagger P_N B_0)^{1/2}.
\end{equation}

Suppose now that  $\{y_1(0),\dots,y_k(0) \}$ is a set of $k$ linearly
independent unit vectors, and let $E_{d \times k}$ denote the $d \times k$ matrix with 1's in the diagonal positions of the $k$ rows, and 0's
elsewhere. The fact that the distribution of $G_i^{(\beta,d)} \vec{x}{}$ is independent of $\vec{x}$ tells us that
\begin{equation}\label{8z}
B_0^\dagger P_N^\dagger P_N B_0 \mathop{=}\limits^{\rm d} 
\prod_{j=1}^N E_{d \times k}^T G_j^{(\beta,d)\, \dagger} \Sigma G_j^{(\beta,d)}  E_{d \times k} =
\prod_{j=1}^N G_{j,k}^{(\beta,d) \, \dagger} \Sigma  G_{j,k}^{(\beta,d)},
\end{equation}
where $G_{j,k}^{(\beta,d)}{}$ denotes the $d \times k$ matrix formed by the first $k$ columns of $G_j^{(\beta,d)}{}$, and it tells us furthermore that the sup operation is
redundant. Consequently
\begin{equation}\label{8}
\mu_1 + \cdots + \mu_k =  \lim_{N \to \infty} \sum_{j=1}^N {1 \over N} \log \det \Big ( G_{j,k}^{(\beta,d) \, \dagger}{} \Sigma  G_{j,k}^{(\beta,d)}{} \Big )^{1/2},
\end{equation}
while the law of large numbers gives that this can be evaluated as the average over $\{G^{(\beta,d)}_k\}$
\begin{equation}\label{8.1a}
\mu_1 + \cdots + \mu_k =   \Big \langle \log \det  \Big ( G_{\cdot,k}^{(\beta,d) \, \dagger}{} \Sigma  G_{\cdot,k}^{(\beta,d)}{} \Big )^{1/2} \Big 
\rangle.
\end{equation}
Computation of this average in the case $\Sigma = \mathbb I_d$ implies \cite{Ne86,Fo13,Ka14}
\begin{equation}\label{33}
\mu_i =    {1 \over 2}  \log {2 \over \beta} + {1 \over 2} \Psi(\beta(d - i + 1)/2).
\end{equation}

Suppose (\ref{1}) is generalised so that each $A_i$ is of size $(d + \nu_i) \times (d + \nu_{i-1})$, $\nu_0 = 0$. It has recently been noted in \cite{Ip14} that the
calculations leading to (\ref{33}) generalise and we have
\begin{equation}\label{33.1}
\mu_i =    {1 \over 2}  \log {2 \over \beta} + {1 \over 2} \langle \Psi(\beta(d - i + 1)/2) \rangle, \quad \langle \Psi(\beta j /2) \rangle := \lim_{N \to \infty}
{1 \over N} \sum_{i=1}^N \Psi ( \beta (\nu_i + j)/2).
\end{equation}
Note that (\ref{33.1}) is a symmetric function of $\{\nu_i\}$. This is in fact a general feature of the joint distribution of the eigenvalues of
$P_N^\dagger P_N$  \cite{IK13}, and so without loss of generality we are 
 free to suppose $0 = \nu_0 \le \nu_1 \le \nu_2 \le \cdots$.
The existence of the limit $\lim_{N \to \infty} \nu_N$ implies that the limit in (\ref{33.1}) is well defined. Hence only a finite number of the matrices $A_i$ can be rectangular, with all the rest being square. In \cite{Ip14} it was furthermore argued, using exact
calculations based on knowledge of the joint distribution of the spectrum of $P_N$ \cite{AB12,ARRS13,Ip13,IK13,Fo14a}
, that the asymptotic form of the eigenvalues $e^{2N y_i}$ in this setting is given by the form (\ref{4}) with $\mu_i$
given by (\ref{33.1}) and
\begin{equation}\label{34}
\sigma_i^2  = {1 \over 4N} \langle \Psi'(\beta(d-i+1)/2) \rangle
\end{equation}
(note that for $\beta = 2$ --- the case of complex entries --- this reproduces the formula for $\sigma_i^2$ in 
(\ref{4.1})).
Our aim in the next subsection is to show how (\ref{34}) can be deduced from analogous calculations as those leading to (\ref{33}).

\subsection{Variances for the standard Gaussian ensembles}\label{S2.2}
Our starting point is the fact that (\ref{7.1}) is valid for finite $N$, provided that the limiting operation is removed on the RHS, and with
$\{ \mu_i \}$ such that the eigenvalues of $P_N^\dagger P_N$ are equal to $\{e^{2N \mu_i}\}$ (see e.g.~the discussion in \cite[\S 6.1]{ABK14}).
In the Gaussian case this simplifies to (\ref{8}) without the limiting operation on the RHS.
The argument leading to (\ref{8.1a}) then gives that to leading order in $N$
\begin{multline}\label{8.1}
\langle (\mu_1 + \cdots + \mu_k)^2 \rangle  - \langle (\mu_1+\cdots+\mu_k) \rangle^2 \\
=   {1 \over N} \bigg ( \Big \langle \bigg ( \log \det  \Big ( G_{\cdot,k}^{(\beta,d) \, \dagger}{} \Sigma  G_{\cdot,k}^{(\beta,d)}{} \Big )^{1/2}  
\bigg )^2 \Big 
\rangle - . \Big \langle \log \det  \Big ( G_{\cdot,k}^{(\beta,d) \, \dagger}{} \Sigma  G_{\cdot,k}^{(\beta,d)}{} \Big )^{1/2}  
 \Big 
\rangle^2  \bigg ).
\end{multline}
The LHS can be rewriten
$$
\sum_{j=1}^k \sigma_j^2 + 2 \sum_{1 \le j < l \le k} ( \langle \mu_j \mu_l \rangle -  \langle \mu_j \rangle \langle \mu_l \rangle ) .
$$
But for large $N$, and a non degenerate Lyapunov spectrum, we expect the covariance $  \langle \mu_j \mu_l \rangle -  \langle \mu_j \rangle \langle \mu_l \rangle $ to decay exponentially fast.
Ignoring such terms, for large $N$ we can substitute
\begin{equation}\label{r}
\sum_{j=1}^k \sigma_j^2 
\end{equation}
for the LHS of (\ref{8.1}). It thus remains to compute the RHS. 
In the case $\Sigma = \mathbb I_k$ this is a straightforward task.

\begin{lemma}\label{L2.1}
We have
\begin{multline}\label{b1}
 \Big \langle \bigg ( \log \det  \Big ( G_{\cdot, k}^{(\beta,d) \, \dagger}{} G_{\cdot,k}^{(\beta,d)}{} \Big )^{1/2} 
\bigg )^2 \Big 
\rangle =  {1 \over 4}  \bigg ( \sum_{j=1}^{k} \Psi'((\beta/2)(d-k) + j\beta/2) \\
+  \Big ( \sum_{j=1}^k  ( \log {2 \over \beta} + \Psi((\beta/2)(d-k) + j\beta/2) ) \Big )^2 \bigg )
\end{multline}
and thus the RHS of (\ref{8.1}) is equal to
\begin{equation}\label{ga}
 {1 \over 4N}   \sum_{j=1}^{k} \Psi'((\beta/2)(d-k) + j\beta/2) .
 \end{equation}
\end{lemma}

\noindent
Proof. \quad We follow the strategy of the proof given in \cite[\S 2.2]{Fo13} of the evaluation of the RHS of (\ref{8.1a}) in the case $\Sigma = \mathbb I_k$ and
$\beta = 2$.

The average in (\ref{b1}) is over $d \times k$  standard Gaussian matrices with real $(\beta = 1)$, complex $(\beta = 2)$ or real quaternion $(\beta = 4)$ entries. Such matrices have probability density function proportional to $e^{-(\beta/2) {\rm Tr} \, ( G_{\cdot,k}^{(\beta,d) \, \dagger}{} G_{\cdot,k}^{(\beta,d)}{})}$, where in the case
$\beta = 4$ the trace operation takes only one the independent eigenvalues (recall the eigenvalues in this case are doubly degenerate). In particular, the
average is a function of $G_{\cdot,k}^{(\beta,d) \, \dagger} G_{\cdot,k}^{(\beta,d)}$, suggesting that we change variables by introducing the $k \times k$ Wishart matrix
$W^{(\beta,k)} = G_{\cdot,k}^{(\beta,d) \, \dagger}{} G_{\cdot,k}^{(\beta,d)}{}$. The corresponding Jacobian is proportional to 
\begin{equation}\label{WG}
(\det W^{(\beta,k)})^{(\beta/2)(d-k+1-2/\beta)}
\end{equation}
(see e.g.~\cite[eq.~(3.22)]{Fo10}). Making use too of the simple identity
$$
{d^2 \over d \mu^2} (\det W^{(\beta,k)})^\mu \Big |_{\mu = 0} =  \Big ( \log \det W^{(\beta,k)} \Big )^2
$$
we see that the LHS of (\ref{b1}) is equal to
\begin{equation}\label{b1a}
{1 \over 4} {d^2 \over d \mu^2} \Big \langle (\det  W^{(\beta,k)})^\mu
\Big \rangle_{W^{(\beta,k)} > 0} \bigg |_{\mu = 0}.
\end{equation}
This average is over positive definite Hermitian matrices with real ($\beta = 1$), complex ($\beta = 2$) or real quaternion ($\beta = 4$) entries, and distribution having a density function
proportional to 
\begin{equation}\label{sat}
(\det  W^{(\beta,k)})^{(\beta/2)(d-k+1-2/\beta) } e^{-(\beta/2) {\rm Tr} \, W^{(\beta,k)}}.
\end{equation}

We see that (\ref{b1a}) is a function only of the eigenvalues of $W^{(\beta,k)}$, suggesting that we change variables to the eigenvalues and the
eigenvectors. The Jacobian --- which has the feature that the eigenvalue and eigenvector terms factorise --- is given in e.g.~\cite[Proposition 1.3.4]{Fo10}, and we deduce that (\ref{b1a}) is equal to
\begin{equation}\label{b1b}
{1 \over 4}  {d^2 \over d\mu^2}   { Z^{(\beta)}_{k,c } \over  Z^{(\beta)}_{k,c_0 }}  \bigg |_{\mu = 0},
\end{equation}
where $c = (\beta/2)(d-k +\mu+1 - 2/\beta)+\mu$, $c_0 =   (\beta/2)(d-k +1 - 2/\beta)$ and
\begin{equation}\label{b1c}
 Z^{(\beta)}_{k,a} := \int_0^\infty dx_1 \cdots \int_0^\infty dx_k \,
 \prod_{j=1}^k e^{-(\beta/2) x_j } x_j^a \prod_{1 \le j < l \le k} |x_l - x_j|^\beta.
 \end{equation}
The multidimensional integral (\ref{b1c}) is familiar in random matrix theory as a limiting case of the Selberg integral, and as such
is evaluated as a product of gamma functions (see e.g.~\cite[Prop.~4.7.3]{Fo10}). This shows that (\ref{b1b}) is equal to
$$
{1 \over 4}  {d^2 \over d \mu^2}  
 \prod_{j=1}^{k} \Big ( {\beta \over 2} \Big )^{-\mu} { \Gamma((\beta/2)(d-k)+\mu +  j \beta/2)  \over
  \Gamma((\beta/2)(d-k)+  j \beta/2) }
 \Big |_{\mu = 0}. 
 $$
 Evaluating the derivative in terms of the digamma function gives (\ref{b1}).
 
 If we now subtract the second term in (\ref{8.1}), recalling (\ref{33}), we see that the term in the
 final line of (\ref{b1}) cancels, leaving (\ref{ga}).
\hfill $\square$
 
\medskip
Equating (\ref{r}) as the evaluation of the LHS of (\ref{8.1}) for large $N$ with the evaluation (\ref{ga}) of the RHS
we obtain
$$
\sum_{j=1}^k \sigma_j^2  =  {1 \over 4N} \ \sum_{j=1}^k \Psi'(\beta(d-j+1)/2) ,
$$
which is equivalent to 
 (\ref{34}) in the case of square matrices.
 
 A minor modification of the above reasoning allows for the derivation of  (\ref{34}) in the case of rectangular matrices. An important feature is that the
 non-negative integer parameters $\nu_i$ ($i=1,\dots,N$) can only take on a finite number of different values. Let these values be $\gamma_1,\dots,\gamma_s$ and
 suppose they are in proportion $\alpha_1,\dots,\alpha_s$ where $\sum_{i=1}^s \alpha_i = 1$. The RHS of the formula (\ref{8.1}) with $\Sigma = \mathbb I_d$ then generalises to 
 read
 \begin{multline}\label{2.18}
{1 \over N}  \bigg ( \sum_{i=1}^s \alpha_i \Big \langle \bigg ( \log \det  \Big ( G_{\cdot,k}^{(\beta,d+\gamma_i) \, \dagger} G_{\cdot,k}^{(\beta,d+\gamma_i)} \Big )^{1/2} 
\bigg )^2 \Big  \rangle\\
-  \sum_{i=1}^s \alpha_i 
 \Big (\Big \langle \log \det  \Big ( G_{\cdot,k}^{(\beta,d+\gamma_i) \, \dagger} G_{\cdot,k}^{(\beta,d+\gamma_i)} \Big )^{1/2}  \Big  \rangle \Big )^2 \bigg ).
\end{multline}
Changing variables $W^{(\beta,k)} = G_{\cdot,k}^{(\beta,d+\gamma_i) \, \dagger} G_{\cdot,k}^{(\beta,d+\gamma_i)}$ for $i=1,\dots,s$ we see that the exponent in the
Wishart ensemble obtained from the corresponding Jacobian is now $  (\beta/2)(d+\gamma_i -k +1 - 2/\beta)$. The result of
Lemma \ref{L2.1} shows that (\ref{b1}) holds with $d$ replaced by $d + \nu$, and all quantities averaged over $\nu$ in the sense of (\ref{33.1}).

A similar averaging of the mean and standard deviation holds in the case that $P_N$ consists of both
Gaussian and inverse Gaussian matrices (see \cite{Fo14} for a study of the singular values in this
setting in the complex case with $N$ finite).  From the definitions, if the Lyapunov exponents for products of
$d \times d$ Gaussian random matrices are $\mu_1^+ > \mu_2^+ > \cdots > \mu_d^+$, then the Lyapunov
exponents for products of $d \times d$ inverse Gaussian random matrices are $- \mu_d^+ > - \mu_{d-1}^+ >
\cdots > - \mu_1^+$. Equivalently, denoting the Lyapunov exponents for products of 
inverse Gaussian matrices by $\mu_1^- > \mu_2^- > \cdots > \mu_d^-$, we thus have
\begin{equation}\label{v1}
\mu_i^- = - \mu_{d+1-i}^+,
\end{equation}
while for the associated variances we have
\begin{equation}\label{v2}
(\sigma_i^-)^2 =  (\sigma_{d+1-i}^+)^2.
\end{equation}

Consider now a matrix product $P_N$ with proportion $\alpha_+$ Gaussian matrices and
$\alpha_-$ inverse Gaussian matrices, and denote the Lyapunov exponents and corresponding variances by $\{\mu_i^{(+,-)}\}$, $\{(\sigma_i^{(+,-)})^2\}$. Consideration of the derivation of (\ref{2.18}), and making use of (\ref{v1})
and (\ref{v2}), we see that
\begin{align}\label{v3}
\mu_i^{(+,-)} &  =  \alpha_+ \mu_i^+ + \alpha_- \mu_i^- =  \alpha_+ \mu_i^+ - \alpha_- \mu_{d+1-i}^+ \nonumber \\
(\sigma_i^{(+,-)} )^2&  =  \alpha_+  
(\sigma_i^+)^2 + \alpha_- (\sigma_i^-)^2 =  \alpha_+  
(\sigma_i^+)^2 + \alpha_- (\sigma_{d+1-i}^+)^2.
\end{align}

\subsection{Complex standard Gaussian ensembles with general $\Sigma$}\label{S2.3}
It has been shown in \cite{Fo13} that for $A_i = \Sigma^{1/2}G_i^{(2,d)}$ with $G_i^{(2,d)}$ a $d \times d$ standard
complex Gaussian random matrix and $\Sigma$ a fixed positive definite matrix with eigenvalues of
$\Sigma^{-1}$ equal to $\{y_1,\dots,y_d\}$, the Lyapunov spectrum corresponding to the product (\ref{1})
is given by
\begin{equation}\label{2.17}
\mu_k = - {1 \over 2 \prod_{1 \le i < j \le d} (y_j - y_i)}
\det \begin{bmatrix} [y_j^{i-1}]_{i=1,\dots,k-1 \atop j =1,\dots,d} \\
[(\log y_j) y_j^{k-1}]_{j=1,\dots,d} \\
[y_j^{i-1}]_{i=k+1,\dots,d \atop j =1,\dots,d} \end{bmatrix} + {1 \over 2} \Psi(k).
\end{equation}
This was possible because of the availability of an explicit formula for the average 
\begin{equation}
\Big \langle \bigg ( \det  \Big ( G_{\cdot,k}^{(2,d) \, \dagger}{} \Sigma  G_{\cdot,k}^{(2,d)}{} \Big )^{1/2} 
\bigg )^\mu \Big 
\rangle.
\end{equation}
Thus, from the works \cite{SMM05,Gh09} we can read off that this average is equal to 
(see e.g.~\cite[eq.~(2.21)]{Fo13})
 \begin{equation}\label{K1}
 \displaystyle
{  (-1)^{k((k+1)/2 - d)} \over \prod_{i=1}^{k-1} i!}
{\prod_{j=1}^d y_j^k \over  \prod_{1 \le j < l \le d} (y_j - y_l)} \
\det \begin{bmatrix} [ \int_0^\infty t^{\nu/2+k-i} e^{-y_j t} \, dt ]_{i=1,\dots,k \atop j=1,\dots, d} \\
[y_j^{d - i}]_{i=k+1,\dots,d \atop j=1,\dots,d} \end{bmatrix}.
\end{equation} 

The Harish-Chandra\slash Itzykson--Zuber matrix integral (see e.g.~\cite[Prop.~11.6.1]{Fo10}), giving a determinant formula for a particular integral over the unitary group is an essential component of the derivation of (\ref{K1}).
The analogous matrix integral over the orthogonal or symplectic group does not permit such a structured evaluation, so impeding further progress for calculating the Lyapunov spectrum of the Gaussian ensembles with general $\Sigma$ and real or real quaternion entries.
Differentiating (\ref{K1}) with respect to $\mu$ and setting $\mu = 0$ gives a formula for the right-hand side
of (\ref{8.1}) in the case $\beta = 2$. This formula is equivalent to (\ref{2.17}).

For purposes of calculating the variances associated with the Lyapunov spectrum, following the strategy
of Section 2.2 we require the analogous evaluation of
 \begin{equation}\label{K2}
 \Big \langle \bigg ( \log \det  \Big ( G_{\cdot, k}^{(2,d) \, \dagger}{} \Sigma G_{\cdot,k}^{(2,d)}{} \Big )^{1/2} 
\bigg )^2 \Big 
\rangle. 
\end{equation}
Moreover, according to (\ref{8.1a}) the second term in (\ref{8.1}) is equal to 
$  ( \sum_{j=1}^k   \langle \mu_j \rangle  )^2 $, and we know from  (\ref{2.17})
the value of $\{\langle \mu_j \rangle\}$. Although the
computation of (\ref{K2}) is elementary, simply requiring a double differentiation of (\ref{K1}) with respect to
$\mu$, unfortunately for general $k$  the
resulting formula is not very revealing, consisting of a number of terms involving either summations
or double summations up to $k$, with some of the summands being determinants. We will
therefore restrict attention to the cases $k=1$, when there is much simplification.
\begin{prop}\label{L2.2}
Let $\{y_1,\dots,y_d\}$ be the eigenvalues of $\Sigma^{-1}$, assumed distinct.
The average (\ref{K2}) in the case $k=1$ is equal to
\begin{equation}\label{b1A}
\mu_1^2 + { \Psi'(1) \over 4} 
 + {1 \over 4}   \sum_{j=1}^d  \displaystyle { (\log y_j)^2 \over \prod_{l=1, l \ne j}^d ( 1 - y_j/y_l)} -
 {1 \over 4} 
\Big (  \sum_{j=1}^d  \displaystyle { \log y_j \over \prod_{l=1, l \ne j}^d ( 1 - y_j/y_l)} \Big )^2 .
\end{equation}
It follows that for large $N$
\begin{equation}\label{b1B}
\sigma_1^2 =  {1 \over 4N} \bigg (
\Psi'(1) 
 +   \sum_{j=1}^d  \displaystyle { (\log y_j)^2 \over \prod_{l=1, l \ne j}^d ( 1 - y_j/y_l)} -
\Big (  \sum_{j=1}^d  \displaystyle { \log y_j \over \prod_{l=1, l \ne j}^d ( 1 - y_j/y_l)} \Big )^2 \bigg ).
\end{equation}
\end{prop}

\noindent
Proof. \quad Our task is to differentiate (\ref{K1}) with $k=1$ twice with respect to 
$\mu$ and to set $\mu = 0$. Since the only dependence on $\nu$ for $k=1$ is in the first row, this reduces to
computing the second derivative of each term in the first row. Noting
$$
{d^2 \over d \mu^2} \int_0^\infty t^{\mu/2} e^{-y_j t} \, dt \Big |_{\mu = 0} =
{1 \over 4 y_j}
\Big ( ( \log y_j)^2 - 2 ( \log y_j) \Psi(1) + \Psi'(1) + (\Psi(1))^2 \Big )
$$
and substituting for the first row of (\ref{K1}) we see after some simple manipulation that the sought average is equal
to
\begin{multline}\label{s.1}
 {1 \over 4 \prod_{1 \le j < l \le d} (y_l - y_j) } \bigg (
\det \begin{bmatrix}
[(\log y_j)^2]_{j=1,\dots,d} \\
[y_j^{i-1}]_{i=2,\dots,d \atop j =1,\dots,d} \end{bmatrix} - 2 \Psi(1)
\det \begin{bmatrix}
[\log y_j]_{j=1,\dots,d} \\
[y_j^{i-1}]_{i=2,\dots,d \atop j =1,\dots,d} \end{bmatrix} \bigg )\\
+ {\Psi'(1) + (\Psi(1))^2 \over 4},
\end{multline}
where in eliminating the determinant which would otherwise appear on the second line use has been
made of the Vandermonde determinant evaluation
\begin{equation}\label{V}
\det [ y_j^{i-1}]_{i=1,\dots,d} = \prod_{1 \le j < l \le d} (y_l - y_j).
\end{equation}
Recalling (\ref{2.17}) in the case $k=1$ allows (\ref{s.1}) to be equivalently written
\begin{multline}\label{s.1}
 \mu_1^2 + {\Psi'(1) \over 4} + {1 \over 4 \prod_{1 \le j < l \le d} (y_l - y_j) } 
\det \begin{bmatrix} 
[(\log y_j)^2]_{j=1,\dots,d} \\
[y_j^{i-1}]_{i=2,\dots,d \atop j =1,\dots,d} \end{bmatrix} \\
- { 1 \over 4 \prod_{1 \le j < l \le d} (y_l - y_j)^2 } 
\bigg (
\det \begin{bmatrix}
[\log y_j]_{j=1,\dots,d} \\
[y_j^{i-1}]_{i=2,\dots,d \atop j =1,\dots,d} \end{bmatrix} \bigg )^2.
\end{multline}
Expanding the determinants by the first row and making use of (\ref{V}) to evaluate the cofactors we obtain
(\ref{b1A}).

To deduce (\ref{b1B}) from (\ref{b1A}) we subtract $\mu_1^2$ in keeping with (\ref{8.1}) and equate with
 (\ref{r}) in the case $k=1$.
\hfill $\square$

\subsection{The variance $\sigma_1^2$ for Gaussian ensembles with general $\Sigma$}\label{S2.4}
Our derivation of the formula (\ref{b1B}) for $\sigma_1^2$ in the case of general variance complex
Gaussian matrices has made use of the determinant formula (\ref{K1}), which as already mentioned
has no analogue in the real or real quaternion cases. This was similarly the case for the derivation given
in \cite{Fo13} of the Lyapunov exponents (\ref{2.17}). However, with regards to the latter with $k=1$
(largest Lyapunov exponent) it has recently been shown by Kargin \cite{Ka14} that an alternative
calculation is possible, which furthermore does generalise to the real $(\beta = 1)$ and real
quaternion $(\beta = 4)$ cases, giving the formula \cite[Th.~1.1]{Ka14}
\begin{equation}\label{Kg}
2 \mu_1 = - \gamma + \log \Big ( {2 \over \beta} \Big ) +
\int_0^\infty \bigg ( \chi_{x \in (0,1)} -
\prod_{i=1}^d \Big ( 1 + {x \over y_i} \Big )^{-\beta/2} \bigg ) \, {dx \over x}.
\end{equation}
In (\ref{Kg}) $\gamma$ denotes Euler's constant and $\chi_J = 1$ for $J$ true and $\chi_J = 0$ otherwise.
Here we adapt the method of \cite{Ka14} to the computation of the RHS of (\ref{8.1}) with $k=1$.

\begin{prop}
Let $\mathcal C$ denote the simple closed contour starting at large $R > {\rm max} \, \{y_i\}$, goes along the upper edge of the real axis to
$\epsilon > 0$, where $\epsilon < {\rm min} \, \{y_i\}$, crosses to the lower edge of the real axis and returns to
$R$. Define
\begin{equation}\label{Jp}
J_p =  {1 \over 2 \pi i}
\int_{\mathcal C} (\log z)^p \prod_{i=1}^d \Big ( 1 - {z \over y_i} \Big )^{-\beta/2} \, {dz \over z} .
\end{equation}
We have
\begin{equation}\label{JpA}
\sigma_1^2 = {1 \over 4 N} \Big ( {\pi^2 \over 6} - J_2 - J_1^2 \Big ).
\end{equation}
\end{prop}

\noindent
Proof. \quad
For $k=1$ the quantity $(G_{\cdot, k}^{(\beta,d) \, \dagger}{} \Sigma G_{\cdot,k}^{(\beta,d)})$ is a positive scalar
random variable. Let its density function be denoted $p_\beta(\lambda)$. We then have that
\begin{equation}\label{21.1}
\Big \langle \Big ( \log \det (G_{\cdot, k}^{(\beta,d) \, \dagger}{} \Sigma G_{\cdot,k}^{(\beta,d)})^{1/2}  \Big )^2 \Big \rangle =
{1 \over 4} \int_0^\infty (\log \lambda )^2 p_\beta(\lambda) \, d \lambda.
\end{equation}
It is shown in \cite{Ka14} that
\begin{equation}\label{21.2}
p_\beta(\lambda) = {1 \over \tilde{c}_\beta}
{1 \over 2 \pi i}
\int_{\mathcal C} e^{-\beta \lambda z /2} \prod_{i=1}^d ( z - y_i)^{-\beta/2} \, dz,
\qquad \tilde{c}_\beta = -{2 \over \beta} \prod_{i=1}^d (-y_i)^{-\beta/2},
\end{equation}
where $\mathcal C$ is as in (\ref{Jp}).
Substituting (\ref{21.2}) in (\ref{21.1}), interchanging the order of integration and noting
$$
\int_0^\infty (\log \lambda)^2 e^{-\beta \lambda z /2} \, d \lambda =
{2 \over \beta z} \bigg (
\Big ( \log {2 \over \beta z} \Big )^2 - 2  \gamma  \Big ( \log {2 \over \beta z} \Big )+ \gamma^2 + {\pi^2 \over 6}
\bigg )
$$
shows that
\begin{equation}\label{21.3}
\int_0^\infty (\log \lambda )^2 p_\beta(\lambda) \, d \lambda. = (\gamma^2 + {\pi^2 \over 6})I_0 -
2 \gamma I_1 + I_2,
\end{equation}
where 
\begin{equation}\label{21.4}
I_p := {1 \over \tilde{c}_\beta} {1 \over 2 \pi i}
\int_{\mathcal C} \Big ( {2 \over \beta z} \Big )
\Big ( \log {2 \over \beta z} \Big )^p \prod_{i=1}^d ( z - y_i)^{-\beta/2} \, dz.
\end{equation}

By deforming the contour to go around the pole at $z=0$ we deduce that $I_0 = 1$.
We read off from \cite[Eq.~(13)]{Ka14} that
\begin{equation}\label{21.4a}
I_1 = 2 \mu_1 + \gamma.
\end{equation}
Also, simple manipulation shows
\begin{align}\label{21.4b}
I_2 & = \Big ( \log {2 \over \beta} \Big )^2 I_0 + 2 \Big  ( \log {2 \over \beta} \Big )(I_1 -\log {2 \over \beta} \Big )
- J_2 \nonumber   \\
& =  \Big ( \log {2 \over \beta} \Big )^2+2 \Big ( \log {2 \over \beta} \Big )\Big ( 2 \mu_1 + \gamma - \log {2 \over \beta}
\Big ) - J_2,
\end{align}
where $J_p$ is specified by (\ref{Jp}). We substitute (\ref{21.4a}) and (\ref{21.4b}) in (\ref{21.3}) then subtract
$(2 \mu_1)^2$, using the fact that \cite[Eq.~(14)]{Ka14}
\begin{equation}\label{21.5}
2 \mu_1 = - \gamma + \log {2 \over \beta} -J_1
\end{equation}
to arrive at (\ref{JpA}). \hfill $\square$

\medskip
In the case $\beta = 2$ the integrals in (\ref{JpA}) can be evaluated using the residue theorem to reclaim (\ref{b1B}), noting too that
$\Psi'(1) = \pi^2/6$. In the case that $y_1 = \cdots = y_d = 1$ and thus $\Sigma = \mathbb I_d$, provided $\beta d/2
\in \mathbb Z^+$ use of the residue theorem shows that
\begin{equation}
-J_1 = \sum_{s=1}^{d\beta/2 - 1} {1 \over s}, \qquad
-J_2 =  \sum_{s=2}^{d\beta/2 - 1} \alpha_s \quad{\rm where} \quad 
\alpha_s = {2 \over s} \sum_{p=1}^{s-1} {1 \over p},
\end{equation}
which when substituted in (\ref{JpA}) can be verified to give the same value of $\sigma_1^2$ as
(\ref{34}) in the case $i=1$.

It is shown in \cite{Ka14} that
\begin{equation}\label{Jz}
J_1 = -\int_0^\infty \bigg ( \chi_{x \in (0,1)} -
\prod_{i=1}^d \Big ( 1 + {x \over y_i} \Big )^{-\beta/2} \bigg ) \, {dx \over x},
\end{equation}
as is consistent with the equality between (\ref{Kg}) and (\ref{21.5}). Analogous working shows
\begin{equation}\label{Jz1}
J_2 = 2 \int_0^\infty \bigg ( \chi_{x \in (0,1)} -
\prod_{i=1}^d \Big ( 1 + {x \over y_i} \Big )^{-\beta/2} \bigg ) {\log x\over x} \, dx + {\pi^2 \over 3},
\end{equation}
so it is also possible to express (\ref{JpA}) in terms of real integrals.

\section{Products of random truncated unitary matrices}\label{S3}
\subsection{Isotropic property}
As remarked in the opening paragraph of \S \ref{S2.1}, the tractability of the computation of the
Lyapunov spectrum for the Gaussian random matrices $G$ is due to Gaussian matrices belonging to the class
of isotropic matrices, meaning that any vector $\vec{x}$, $G \vec{x}/|\vec{x}|$ has distribution
independent of $\vec{x}$. This property holds for any class of random matrices that are unitary invariant, and
thus their distribution in unchanged by multiplication by a unitary matrix $U$, which is furthermore required to
have entries of the same type (real, complex or real quaternion) as the random matrices.

Let $Z^{(\beta,d+n)}$ be a random unitary matrix of size $(d+n) \times (d+n)$ with real ($\beta = 1$,
complex ($\beta = 2$) or real quaternion ($\beta = 4$) elements, and chosen with Haar measure. The latter
 has the defining feature that it is unchanged by multiplication of the matrices by fixed unitary matrices
of the appropriate type. Suppose the fixed unitary matrices have  the block structure
$$
U = \begin{bmatrix} V_{d \times d} & \\
 & \mathbb I_n \end{bmatrix},
 $$
 where $V_{d \times d}$ is a $d \times d$ unitary matrix. Such matrices act only on the top $d \times d$
 sub-matrix  of $Z^{(\beta,d+n)}$, so it must be that square submatrices of Haar distributed unitary matrices
 (also referred to as truncated unitary matrices \cite{ZS99}) are themselves unitary invariant, and are thus examples of isotropic matrices. With the matrices $A_i$ in
 (\ref{1}) chosen from the ensemble of truncated unitary matrices, our interest in this section is to compute
 the Lyapunov spectrum as well as the associated variances. Crucial to our computations is the fact that
 a $d \times k$ ($d \ge k$) sub-block $Z^{(\beta,d) }_{\cdot,k}$ is distributed with a probability density
 function proportional to 
 \begin{equation}\label{Z}
 \Big ( \det (\mathbb I_k - Z^{(\beta,d) \, \dagger}_{\cdot,k} Z^{(\beta,d) }_{\cdot,k} )\Big )^{c_2}, \qquad
 c_2 := (\beta/2)(n - k + 1 - 2/\beta).
 \end{equation}
 
 We remark that the singular values of products of truncated complex unitary matrices for finite $N$
 are the topic of the recent work \cite{KKS15}.
 \subsection{Computation of the Lyapunov spectrum and variances}
 According to (\ref{8.1a}) we have
\begin{equation}\label{8.1Z}
\mu_1 + \cdots + \mu_k =   \Big \langle \log \det  \Big ( Z^{(\beta,d) \, \dagger}_{\cdot,k} Z^{(\beta,d) }_{\cdot,k}  \Big )^{1/2} \Big 
\rangle,
\end{equation}
where the average is with respect to (\ref{Z}). The RHS of (\ref{8.1Z}) is easy to compute, allowing for a simple
formula for the individual $\mu_i$.

\begin{prop}\label{PU}
Let the matrices $A_i$ in (\ref{1}) be chosen as the top $d \times d$ sub-block of a $(d + n) \times
(d + n)$ Haar distributed random unitary matrix with real $(\beta = 1)$, complex $(\beta = 2)$ or
real quaternion $(\beta = 4)$ entries. For $n \ge d$ at least we have
\begin{equation}\label{arr}
\mu_i =  {1 \over 2} \Big ( \Psi((\beta/2)(d-i+1)) - \Psi((\beta/2)(n+d-i+1) \Big ).
\end{equation}
\end{prop}

\noindent
Proof. \quad As in the Gaussian case, an essential point is that the average (\ref{8.1Z}) is a function only of the
positive definite matrix $W^{(\beta,k)} =  Z^{(\beta,d) \, \dagger}_{\cdot,k} Z^{(\beta,d) }_{\cdot,k} $.
Since $Z^{(\beta,d) }_{\cdot,k} $ is a sub-matrix of a unitary matrix, $W^{(\beta,k)}$ is further constrained
to have eigenvalues less than 1, which we denote by writing $W^{(\beta,k)} < 1$. Regarding this as a change
of variables, the corresponding Jacobian is proportional to (\ref{WG}). Hence our task becomes that of
evaluating
\begin{equation}\label{14a}
{1 \over 2} {d \over d \mu}  \Big \langle (\det W^{(\beta,k)} )^\mu \Big \rangle_{0 < W^{(\beta,k)} < 1}
\Big |_{\mu = 0},
\end{equation}
where $W^{(\beta,k)}$ has distribution with a probability density function proportional to
\begin{equation}\label{14b}
(\det W^{(\beta,k)} )^{c_1}  \Big ( \det (\mathbb I_k -  W^{(\beta,k)} \Big )^{c_2} 
\end{equation}
with $c_2$ as in (\ref{Z}) and
\begin{equation}\label{c1}
c_1 =  (\beta/2)( d - k + 1 - 2/\beta).
\end{equation}

Both (\ref{14a}) and (\ref{14b}) are functions of the eigenvalues of $ W^{(\beta,k)}$ only. Introducing the
appropriate Jacobian as can be found in e.g.~\cite[Proposition 1.3.4]{Fo10} gives for (\ref{14a}) the 
expression in terms of multiple integrals
\begin{equation}\label{14c}
{1 \over 2} {d \over d \mu} {I_{k,c_1+\mu,c_2}^{(\beta)} \over I_{k,c_1,c_2}^{(\beta}} \bigg |_{\mu = 0},
\end{equation}
where
\begin{equation}\label{5.1}
 I_{k,\alpha_1,\alpha_2}^{(\beta)} =
 \int_0^1 dx_1 \cdots \int_0^1 dx_k \, \prod_{l=1}^k x_l^{\alpha_1} (1 - x_l)^{\alpha_2}
 \prod_{1 \le j < l \le k} |x_l - x_j|^\beta.
 \end{equation}
 This is precisely the Selberg integral (see e.g.~\cite[Ch.~4]{Fo10}), already seen in a limiting form in
 (\ref{b1c}). Inserting its gamma functions evaluation reduces (\ref{14c}) to
 \begin{multline}\label{19a}
 {1 \over 2} {d \over d \mu}
 \prod_{j=0}^{k-1}
 {\Gamma(c_1+\mu+1+j \beta/2) 
 \Gamma(c_1+c_1+c_2+2+(k+j-1)\beta/2) \over
 \Gamma(c_1+1+j \beta/2) 
 \Gamma(c_1+\mu+c_2+2+(k+j-1)\beta/2)} \bigg |_{\mu=0} \\
  = {1 \over 2} \Big (
 \sum_{j=0}^{k-1} \Psi(c_1 + 1 + j\beta/2) - \sum_{j=0}^{k-1}
 \Psi(c_1+c_2+2 + (k+j-1)\beta/2) \Big ).
 \end{multline}
 Equating this with the LHS of (\ref{8.1Z}) and inserting the values of $c_1$ and $c_2$ from (\ref{c1}) and
 (\ref{Z}) we arrive at (\ref{arr}). \hfill $\square$
 
 \medskip
 We remark that even though our derivation of (\ref{arr}) requires $n \ge d$, used for example in the convergence of
the integrals in (\ref{14c}), the final expression is well defined for $n \ge 0$, and we expect it to remain valid.
Note in particular that for $n=0$, (\ref{arr}) gives that each Lyapunov is equal to 0. This is consistent
with the matrices in the product (\ref{1}) then  themselves being unitary matrices, as the $d \times d$
sub-matrix of
$Z^{(\beta,d+n)}$ is for $n=0$ the
matrix itself.

Similar working gives the variances associated with the Lyapunov exponents (\ref{arr}).

\begin{prop}\label{PL}
The analogue of the formula (\ref{34}) for the variances associated with the Lyapunov exponents in the
case of products of the truncated unitary matrices of Proposition \ref{PU} is given by
\begin{equation}\label{arr1}
\sigma_i^2 =  {1 \over 4N} \Big ( \Psi'((\beta/2)(d-i+1)) - \Psi'((\beta/2)(n+d-i+1) \Big ).
\end{equation}
\end{prop}

\noindent
Proof. \quad We use the analogue of the formula (\ref{8.1}) with the LHS replaced by its leading large $N$ form (\ref{r}).
Thus our main task is to compute
\begin{equation}
\Big \langle \Big (  \log \det  \Big ( Z^{(\beta,d) \, \dagger}_{\cdot,k} Z^{(\beta,d) }_{\cdot,k}  \Big )^{1/2} 
\Big )^2 \Big \rangle.
\end{equation}
Following the working in the proof of Proposition \ref{PU} we see that this average is equal to
\begin{align*}\label{14cq}
&{1 \over 4} {d^2 \over d \mu^2} {I_{k,c_1+\mu,c_2}^{(\beta)} \over I_{k,c_1,c_2}^{(\beta}} \bigg |_{\mu = 0} \nonumber \\
&=
{1 \over 4} {d^2 \over d \mu^2}
 \prod_{j=0}^{k-1}
 {\Gamma(c_1+\mu+1+j \beta/2) 
 \Gamma(c_1+c_1+c_2+2+(k+j-1)\beta/2) \over
 \Gamma(c_1+1+j \beta/2) 
 \Gamma(c_1+\mu+c_2+2+(k+j-1)\beta/2)} \bigg |_{\mu=0} \\
 & = {1 \over 4} \Big (
 \sum_{j=0}^{k-1} \Psi'(c_1 + 1 + j\beta/2) - \sum_{j=0}^{k-1}
 \Psi'(c_1+c_2+2 + (k+j-1)\beta/2) \Big ) \\
 & \qquad + {1 \over 4} \Big (
 \sum_{j=0}^{k-1} \Psi(c_1 + 1 + j\beta/2) - \sum_{j=0}^{k-1}
 \Psi(c_1+c_2+2 + (k+j-1)\beta/2) \Big )^2.
\end{align*}
We recognise from (\ref{19a}) that the final line is equal to $(\sum_{i=1}^d \mu_i)^2$, and so cancels out of
the RHS of (\ref{8.1}).
Thus
$$
\sum_{i=1}^k \sigma_i^2 = {1 \over 4} \Big (
 \sum_{j=0}^{k-1} \Psi'(c_1 + 1 + j\beta/2) - \sum_{j=0}^{k-1}
 \Psi'(c_1+c_2+2 + (k+j-1)\beta/2) \Big ).
 $$
Recalling the values of $c_1$ and $c_2$ from (\ref{c1}) and
 (\ref{Z}) we see that this is equivalent to the stated formula (\ref{arr1}). \hfill $\square$
 
 \medskip

The formulas (\ref{arr}) and (\ref{arr1}) have some similarity with the corresponding formulas
(\ref{33.1}) and  (\ref{34}) for products of rectangular Gaussian random matrices, one of size
$d \times d$ (and thus square), the other of size $(d+n) \times d$. In particular the formulas agree if
the $d \times d$ matrices were to occur in ``proportion" 1, and the $(d+n) \times d$ matrices in ``proportion" $-1$.
  
  \section{Discussion}
  \subsection{Comparison with numerical experiments}
  In our earlier paper on Lyapunov exponents \cite{Fo13} a discussion and references
  were given on the difficulty of accurately computing the Lyapunov exponents.
  A more robust, but less accurate approach is to use Monte Carlo simulation based on (\ref{7.1}).
  Numerically stable ways  to carry out such computations have been reported in \cite{CPV93}.
  Thus, in relation to $\mu_1$, and with $d=2$, $\vec{x}_0 = \begin{bmatrix}1 & 0 \end{bmatrix}^T$,
  we iteratively define $\vec{x}_i = A_i \vec{y}_{i-1}$ $(i=1,2,\dots)$, where $\vec{y}_i =
  \vec{x}_i/|\vec{x}_i|$. According to (\ref{7.1}), a Monte Carlo estimate of the largest Lyapunov
  exponent is then given by
  \begin{equation}\label{Mon}
  \mu_1 = {1 \over N} \sum_{j=1}^N \log |\vec{x}_j|,
  \end{equation}
  and similarly an estimate of the corresponding variance is given by
  \begin{equation}\label{Mon1}
  \sigma_1^2 = {1 \over N} \Big (  {1 \over N} \sum_{j=1}^N (\log |\vec{x}_j|)^2 -
  \mu_1^2 \Big ).
  \end{equation}
  To use (\ref{7.1}) to compute $\mu_1 + \mu_2$, again taking $d=2$ for simplicity, let 
  $\vec{x}_0^{(1)} = \begin{bmatrix}1 & 0 \end{bmatrix}^T$ and 
  $\vec{x}_0^{(2)} = \begin{bmatrix}0 & 1 \end{bmatrix}^T$, then define
  $\vec{x}_i^{(p)} = A_i \vec{y}_i^{(p)}$, ($p=1,2$, $i=1,2,\dots$) where $\{\vec{y}_i^{(1)}, \vec{y}_i^{(2)} \}$ is obtained from
 $\{\vec{x}_{i-1}^{(1)}, \vec{x}_{i-1}^{(2)} \}$ by the Gram-Schmidt orthonormalization procedure. It follows
 from (\ref{7.1}) that a Monte Carlo estimate of $\mu_1 + \mu_2$ is
  \begin{equation}\label{Mon2}
\mu_1 +  \mu_2 =    {1 \over N} \sum_{j=1}^N \log \det (X_j^\dagger X_j)^{1/2},
 \end{equation}
 where $X_i$ is the $2 \times 2$ matrix with columns given by $\vec{x}_i^{(1)}$ and  $\vec{x}_i^{(2)}$, while an
 estimate of the corresponding variance is
  \begin{equation}\label{Mon3}
  \sigma_1^2 + \sigma_2^2 = {1 \over N} \Big ( {1 \over N} \sum_{j=1}^N (\log \det (X_j^\dagger X_j)^{1/2})^2 -
  (\mu_1 + \mu_2)^2 \Big ) .
  \end{equation}
  
  As an example, let's consider the case of (\ref{1}) with $A_i = \Sigma^{1/2} G_i^{(2,2)}$. We choose the
  eigenvalues
  of $\Sigma^{-1}$ to equal $1,1/4$ so that
  $$
  A_i = \begin{bmatrix} 1 & 0 \\ 0 & 2 \end{bmatrix} G_i^{(2,2)}.
  $$
  The formula (\ref{b1B}) gives $N \sigma_1^2 = \Big ( {\pi^2 \over 24} - {4 \over 9} ( \log 2)^2 \Big ) =
  0.197699...$, whereas a Monte Carlo calculation based on (\ref{Mon1}) with $N = 10^6$ gave
  $N \sigma_1^2 \approx 0.19686$.
  
  To perform Monte Carlo simulations in relation to products of truncated unitary matrices, we must first generate
  random unitary matrices with Haar measure. One way to do this is to apply the Gram-Schmidt orthogonalization 
  producure to a standard Gaussian matrix. In the complex case, with $d=n=2$ we did this and with $N = 10^5$
  found the estimates $\mu_1 \approx -0.41672$ and $\mu_1 + \mu_2 \approx -1.6685$ whereas according to
  Proposition \ref{PU} $\mu_1 = - {5 \over 12} =-0.4166...$ and $\mu_1 + \mu_2 = - {7 \over 6} = -1.666...$.
  For the corresponding variances we found $N \sigma_1^2 \approx 0.0894$ and
  $N(\sigma_1^2 + \sigma_2^2) \approx 0.3952$ whereas Proposition \ref{PL} gives
  $N\sigma_1^2 = 13/144 = 0.0902...$ and $N(\sigma_1^2 + \sigma_2^2) =  29/72 \approx 0.4027...$.

  \subsection{Asymptotic spectrum of $P_N$}\label{S4.2}
  As revised in the first paragraph of the Introduction, the Lyapunov exponents are defined in terms of
   the eigenvalues of the
  positive definite matrix $P_N^\dagger P_N$, where $P_N$ is the random matrix product (\ref{1}). In the recent
  works \cite{ABK14,Ip14}, for the particular cases that the $A_i$ in (\ref{1}) are standard Gaussian matrices with
  real, complex or real quaternion entries, explicit asymptotic analysis of the joint distribution of the spectrum
  of $P_N$ has revealed that with the eigenvalues written in the form $z_k = e^{N \lambda_k + i \theta_k}$, the
  $\{\lambda_k \}$ --- referred to as the stability exponents --- are identical to the Lyapunov exponents.
  Moreover in \cite{ABK14} a proof of the equality between the Lyapunov and stability exponents in the case
  of $2 \times 2$ random matrices from an isotropic ensemble has been given. It was then conjectured that all
  matrix products formed from an isotropic ensemble have this property.
  
  Actually the question of the asymptotic equality between the Lyapunov and stability exponents has been
  investigated at least as far back as 1987 \cite{GSO87}, albeit in the setting of the linearization of sets
  of differential equations about fixed points. In this work it was conjectured that the exponents agree whenever
  the spectrum of $P_N$ is non-degenerate. The reference book \cite{CPV93} cites \cite{GSO87}, and furthermore
  addresses directly the case of products of random matrices, writing ``If the matrix $P_N$ can be written in
  diagonal form, then [in our notation] $\lambda_k = \mu_k$". One immediate consequence of this conjecture
  is the prediction of the Lyapunov spectrum for a product of upper triangular random matrices. There, up
  to ordering $\{\lambda_k\} = \{ \log \langle |A_{kk} | \rangle \}$ where $A_{kk}$ denotes the entry $(kk)$ in the matrices
  $A_i$ of (\ref{1}), so the conjecture gives this same set for the
  Lyapunov exponents. In the case of $\mu_1$ this agrees with a known theorem \cite{Pi85}.
  
  Following \cite{GJJN03}, we emphasize that the conjecture $\lambda_k = \mu_k$ appears to be an
  intrinsic property of the asymptotic limit of random matrix products, and is not shared by other asymptotic
  limits of spectra in random matrix theory. For example, in the case of $N \times N$ standard real,
  complex or real quaternion standard Gaussian random matrices
  $$
  \lim_{N \to \infty} {\nu_1((X^\dagger X)^{1/2}) \over |\nu_1(X)|} = \sqrt{2},
  $$
  where $\nu_1(Y)$ denotes the eigenvalue of largest modulus of $Y$. That this ratio is at least unity follows
  from the general inequality $|\nu_1(Y)| \le \nu_1((Y^\dagger Y)^{1/2})$, which in turn is a consequence of the
  sub-multiplicity of the RHS viewed as a matrix norm.
  
\subsection*{Acknowledgements}
 This work was supported by the Australian Research Council for the project DP140102613.
 I thank Jesper Ipsen for comments on the first draft.
 

\providecommand{\bysame}{\leavevmode\hbox to3em{\hrulefill}\thinspace}
\providecommand{\MR}{\relax\ifhmode\unskip\space\fi MR }
\providecommand{\MRhref}[2]{%
  \href{http://www.ams.org/mathscinet-getitem?mr=#1}{#2}
}
\providecommand{\href}[2]{#2}

\end{document}